\documentstyle[preprint,eqsecnum,epsf,aps,floats,tighten]{revtex}
\input epsf

\newcommand{\XX}{\mbox{$\, \times \,$}}
\newcommand{\pbarp}{\mbox{$\bar{p}p$}}
\newcommand{\pbar}{\mbox{$\bar{p}$}}

\def\gevcc{GeV/$c^2$}                   %GeV/c^2

\newcommand{\cmehi}{\mbox{${\sqrt{s}=1800}$\,GeV}}
\newcommand{\cmelo}{\mbox{$\sqrt{s}=630$\,GeV}}

\newcommand{\rs}{\mbox{$\sqrt{s}$}}
\newcommand{\ncal}{\mbox{$n_{{\rm CAL}}$}}
\newcommand{\nl}{\mbox{$n_{{\rm L\O}}$}}
\def\met{\mbox{${\hbox{$E$\kern-0.6em\lower-.1ex\hbox{/}}}_T$}} %missing ET
\def\gevcc{GeV/$c^2$}                   %GeV/c^2

\draft
\begin{document}

\title{
Observation of Diffractively Produced $W$ and $Z$ Bosons in
$\bbox{\bar{p}p}$ Collisions at $\bbox{\sqrt{s}}$ =  1800 GeV }

% LIST_OF_AUTHORS_R1.TEX                 7/24/03
%
\author{
%% names begin here
V.M.~Abazov,$^{21}$ B.~Abbott,$^{55}$ A.~Abdesselam,$^{11}$
M.~Abolins,$^{48}$ V.~Abramov,$^{24}$ B.S.~Acharya,$^{17}$
D.L.~Adams,$^{53}$ M.~Adams,$^{35}$ S.N.~Ahmed,$^{20}$
G.D.~Alexeev,$^{21}$
% ...
\\
\vskip 0.30cm
\centerline{(D\O\ Collaboration)}
\vskip 0.30cm
}
\address{
\centerline{$^{1}$Universidad de Buenos Aires, Buenos Aires, Argentina}
\centerline{$^{2}$LAFEX, Centro Brasileiro de Pesquisas F{\'\i}sicas,
                  Rio de Janeiro, Brazil}
\centerline{$^{3}$Universidade do Estado do Rio de Janeiro,
                  Rio de Janeiro, Brazil}
\centerline{$^{4}$Institute of High Energy Physics, Beijing,
                  People's Republic of China}
\centerline{$^{5}$Universidad de los Andes, Bogot\'{a}, Colombia}
\centerline{$^{6}$Charles University, Center for Particle Physics,
                  Prague, Czech Republic}
\centerline{$^{7}$Institute of Physics, Academy of Sciences, Center
                  for Particle Physics, Prague, Czech Republic}
\centerline{$^{8}$Universidad San Francisco de Quito, Quito, Ecuador}
\centerline{$^{9}$Laboratoire de Physique Subatomique et de Cosmologie,
                  IN2P3-CNRS, Universite de Grenoble 1, Grenoble, France}
\centerline{$^{10}$CPPM, IN2P3-CNRS, Universit\'e de la M\'editerran\'ee,
                  Marseille, France}
\centerline{$^{11}$Laboratoire de l'Acc\'el\'erateur Lin\'eaire,
                  IN2P3-CNRS, Orsay, France}
\centerline{$^{12}$LPNHE, Universit\'es Paris VI and VII, IN2P3-CNRS,
                  Paris, France}
\centerline{$^{13}$DAPNIA/Service de Physique des Particules, CEA, Saclay,
                  France}
\centerline{$^{14}$Universit{\"a}t Mainz, Institut f{\"u}r Physik,
                  Mainz, Germany}
\centerline{$^{15}$Panjab University, Chandigarh, India}
\centerline{$^{16}$Delhi University, Delhi, India}
\centerline{$^{17}$Tata Institute of Fundamental Research, Mumbai, India}
\centerline{$^{18}$CINVESTAV, Mexico City, Mexico}
\centerline{$^{19}$FOM-Institute NIKHEF and University of
                  Amsterdam/NIKHEF, Amsterdam, The Netherlands}
\centerline{$^{20}$University of Nijmegen/NIKHEF, Nijmegen, The
                  Netherlands}
\centerline{$^{21}$Joint Institute for Nuclear Research, Dubna, Russia}
\centerline{$^{22}$Institute for Theoretical and Experimental Physics,
                   Moscow, Russia}
\centerline{$^{23}$Moscow State University, Moscow, Russia}
\centerline{$^{24}$Institute for High Energy Physics, Protvino, Russia}
\centerline{$^{25}$Lancaster University, Lancaster, United Kingdom}
\centerline{$^{26}$Imperial College, London, United Kingdom}
\centerline{$^{27}$University of Arizona, Tucson, Arizona 85721}
\centerline{$^{28}$Lawrence Berkeley National Laboratory and University of
                  California, Berkeley, California 94720}
\centerline{$^{29}$University of California, Davis, California 95616}
\centerline{$^{30}$California State University, Fresno, California 93740}
\centerline{$^{31}$University of California, Irvine, California 92697}
\centerline{$^{32}$University of California, Riverside, California 92521}
\centerline{$^{33}$Florida State University, Tallahassee, Florida 32306}
\centerline{$^{34}$Fermi National Accelerator Laboratory, Batavia,
                   Illinois 60510}
\centerline{$^{35}$University of Illinois at Chicago, Chicago,
                   Illinois 60607}
\centerline{$^{36}$Northern Illinois University, DeKalb, Illinois 60115}
\centerline{$^{37}$Northwestern University, Evanston, Illinois 60208}
\centerline{$^{38}$Indiana University, Bloomington, Indiana 47405}
\centerline{$^{39}$University of Notre Dame, Notre Dame, Indiana 46556}
\centerline{$^{40}$Iowa State University, Ames, Iowa 50011}
\centerline{$^{41}$University of Kansas, Lawrence, Kansas 66045}
\centerline{$^{42}$Kansas State University, Manhattan, Kansas 66506}
\centerline{$^{43}$Louisiana Tech University, Ruston, Louisiana 71272}
\centerline{$^{44}$University of Maryland, College Park, Maryland 20742}
\centerline{$^{45}$Boston University, Boston, Massachusetts 02215}
\centerline{$^{46}$Northeastern University, Boston, Massachusetts 02115}
\centerline{$^{47}$University of Michigan, Ann Arbor, Michigan 48109}
\centerline{$^{48}$Michigan State University, East Lansing, Michigan 48824}
\centerline{$^{49}$University of Nebraska, Lincoln, Nebraska 68588}
\centerline{$^{50}$Columbia University, New York, New York 10027}
\centerline{$^{51}$University of Rochester, Rochester, New York 14627}
\centerline{$^{52}$State University of New York, Stony Brook,
                   New York 11794}
\centerline{$^{53}$Brookhaven National Laboratory, Upton, New York 11973}
\centerline{$^{54}$Langston University, Langston, Oklahoma 73050}
\centerline{$^{55}$University of Oklahoma, Norman, Oklahoma 73019}
\centerline{$^{56}$Brown University, Providence, Rhode Island 02912}
\centerline{$^{57}$University of Texas, Arlington, Texas 76019}
\centerline{$^{58}$Rice University, Houston, Texas 77005}
\centerline{$^{59}$University of Virginia, Charlottesville, Virginia 22901}
\centerline{$^{60}$University of Washington, Seattle, Washington 98195}
}
%end

\date{\today}
\maketitle
%pre add
%take this out now for pre due to longer author list \pagebreak
\vspace{1cm}
%agb
\setcounter{section}{1}

%agb for pre add pagebreak
\pagebreak

\begin{abstract}
Using the D\O\ detector, we have observed events produced in
\pbarp\ collisions that contain $W$ or $Z$ bosons in conjunction
with very little energy deposition (``rapidity gaps'') in large
forward regions of the detector. The fraction of $W$ boson events
with a rapidity gap (a signature for diffraction) is
$0.89\pm^{0.19}_{0.17}\%$, and the probability that the
non-diffractive background fluctuated to yield the observed
diffractive signal is $3 \times 10^{-14}$, corresponding to a
significance of 7.5$\sigma$. The $Z$ boson sample has a gap
fraction of $1.44 \pm^{0.61}_{0.52}\%$, with a significance of
4.4$\sigma$. The diffractive events have very similar properties
to the more common non-diffractive component.
%The
%fraction of events with these vector bosons and an associated
%rapidity gap is measured, along with the Diffractive $W$/$Z$-boson
%cross section ratio. Event properties for  the Diffractive
%$W$-boson sample are compared to the non-diffractive sample.

\end{abstract}

%\pacs{12.38.Qk, 12.40 Nn, 14.70 Fm }

%pre delete
%\twocolumn
%\pagebreak

\vspace{1.0cm} Inelastic diffractive collisions are responsible
for about 15\% of the \pbarp\ total cross section, and have been
described by Regge theory through the exchange of a
pomeron~\cite{regge}. Such events are characterized by a proton
(or antiproton) carrying away most of the beam momentum, and by
the absence of significant hadronic particle activity over a large
region of pseudorapidity ($\eta=-\ln[\tan(\frac{\theta}{2})]$,
where $\theta$ is the polar angle relative to the beam). This
empty region is called a rapidity gap and can be used as an
experimental signature for diffraction. Ingelman and Schlein
proposed the possibility of a partonic structure for the pomeron,
which would lead to hard scattering in diffractive
events~\cite{IS}.  This so-called ``hard diffraction'' was first
observed by the UA8 experiment~\cite{UA8} at the CERN $Sp \pbar S$
collider in the form of jet events with an associated tagged
proton.

Initial rapidity-gap-based analyses of diffractive
jet~\cite{CDFJ,Zeus,H1}, $b$-quark~\cite{CDFb}, and
$J/\Psi$~\cite{CDFjpsi} production are qualitatively consistent
with a predominantly hard gluonic pomeron, but the production
cross sections observed at the Fermilab Tevatron are far lower
than predictions based on data from the DESY $ep$ collider
HERA~\cite{CDFJ,fact}. Diffractive jet results from the CDF
collaboration using an antiproton tag~\cite{CDFpbj} confirm the
normalization discrepancy between Tevatron (\cmehi) and HERA data,
while recent D\O\ rapidity-gap-based diffractive jet results at
\cmehi\
%agb check spacing after cmelo
and \cmelo~\cite{D0jetPLB} show that a simple normalization
difference cannot accommodate the Tevatron data (and imply that a
significant soft gluon component is needed to ``save'' the
Ingelman-Schlein model).  A unified picture within this framework
requires a detailed understanding of gap survival probability,
which includes effects from multiple parton scattering and extra
gluon emission associated with the hard sub-process~\cite{khoze}.
The soft color interaction (SCI) model~\cite{SCI}, which
hypothesizes that non-perturbative gluon emissions can create
rapidity gaps, provides an alternative description of diffraction
without invoking pomeron dynamics, and predicts diffractive rates
similar to those observed.
%Experimental studies of hard single
%diffraction (HSD), which combines diffraction and a hard scatter
%(such as jet or $W$-boson production), can be used to determine
%the properties of the pomeron, or whether pomeron-based models are
%useful.

Bruni and Ingelman~\cite{bruni} proposed that a search for
diffractive production of $W$ and $Z$ bosons would provide
important information on diffractive structure, due to their
sensitivity to quark sub-structure. They predicted that a pomeron
composed primarily of quarks would lead to more than 15\% of $W$
and $Z$ bosons being diffractively produced. The SCI model, on the
other hand, predicts a diffractive fraction of about
1\%~\cite{SCI2}.

The CDF collaboration observed a 3.8 standard deviation ($\sigma$)
signal for diffractive $W$ boson production, extracting the signal
using the asymmetry of both  lepton charge and position relative
to the region of the rapidity gap, and obtained a diffractive to
non-diffractive production ratio of $(1.15\pm
0.55)$\%~\cite{CDFW2}. In this Letter, we present a definitive
observation of diffractively produced $W$ and $Z$ bosons. We
present characteristics of diffractive $W$ bosons, and
measurements of the fraction of $W$ and $Z$ boson events that
contain forward rapidity gaps. In addition, we provide the ratio
of  diffractive $W$ and $Z$ cross sections, and the fraction of
the initial momentum carried away by the scattered proton in the
collision.
%These unique measurements place significant new constraints on the
%pomeron and diffractive models.

In the D\O\ detector~\cite{NIM},
%agb remove jets if don't include mc jet info
%jets,
electrons are measured and missing transverse energy (\met)
determined using the uranium/liquid-argon calorimeters, with
electromagnetic coverage to $|\eta|\!=\!4.1$ and coverage for
hadrons to $|\eta|\!=\!5.2$. Electron identification, described in
more detail below, requires a central or forward drift chamber
track to match the location of the associated electromagnetic
cluster. For the period during which the data were collected, the
D\O\ detector had no magnetic field within the central tracking
volume, consequently electrons and positrons could not be
differentiated and are both referred to as electrons.
%agb remove jets and d0jet ref
%Jets are reconstructed using a fixed-cone algorithm with radius
%${\cal R} = \sqrt{(\Delta\eta)^2 + (\Delta\phi^2)}=0.7$ ($\phi$ is
%the azimuthal angle). The jet energy is corrected as described in
%Ref.~\cite{D0jets}.

To identify rapidity gaps, we count the number of tiles containing
a signal in the L\O\ forward scintillator arrays (\nl) and the
number of calorimeter towers ($\Delta\eta \XX \Delta\phi = 0.1 \XX
0.1$) with signals above threshold (\ncal). The L\O\ arrays
provide partial coverage in the region $2.3<|\eta|<4.3$. A portion
of the two forward calorimeters ($3.0<|\eta|<5.2$) is used to
measure the calorimeter multiplicity, with a particle tagged by
the deposition of more than $150$ (500)\,MeV of energy in an
electromagnetic (hadronic) tower. These thresholds are set to
minimize noise from  radioactive decays in the uranium, while
maximizing sensitivity to energetic particles~\cite{Kristal}.

For this analysis, we search for the presence of rapidity gaps in
inclusive samples of $W \rightarrow e\nu$ events and  $Z
\rightarrow e^+e^-$ events, based on data with an integrated
luminosity of approximately 85 pb$^{-1}$ accumulated during the
1994--1995 collider run (Run Ib). The D\O\ collaboration has
extensively studied $W$ and $Z$ boson production in the electron
channel~\cite{wmass_prd,wwidth_prd}. The requirements for the
event selection in this analysis are nearly identical to those of
Ref.~\cite{wwidth_prd}, with two notable exceptions detailed
below.  The data were obtained using a single hardware trigger
that required at least one electromagnetic (EM) object with
transverse energy ($E_T$) greater than 15 GeV, with more than 85\%
of its energy deposited in the EM section of the calorimeter (EM
fraction). At the software trigger level, the EM cluster is
required to satisfy isolation, shower-shape, and EM fraction
criteria consistent with the presence of an electron. For the $W$
boson sample, we require this candidate electron to have an $E_T>$
20 GeV, and additionally require $\met
>  15$ GeV for the neutrino, while for the $Z$ boson sample, we require
two electron candidates with $E_T>16$ GeV.

The first significant difference between the data samples in this
analysis and those of Ref.~\cite{wwidth_prd} is that we are unable
to include events from the first portion of Run Ib, during which a
coincidence (in the L\O\ detector) between the remnants of the
proton and antiproton was required, effectively vetoing
single-diffractive production. Restricting this analysis to the
part of the data collected without this condition reduces the
sample by 30\%. The only other major difference is that this
analysis requires the removal of events with more than one
proton-antiproton interaction in the same bunch crossing. This
``single interaction'' requirement is necessary for
rapidity-gap-based diffractive studies, because the presence of
additional events obscures the rapidity-gap signature. About 70\%
of the remaining data sample is discarded as a result of this
requirement, which makes use primarily of timing information in
the luminosity counters and the number of vertices found in the
central tracker to reject multiple interaction events.

%agb include invariant mass comments--inv mass window in table
%%%%%%%%%%%%%%%%%%%%%%%%%%%%%%%%%%%%
%%%%%%%%%%%%%%%%%%%%%%%%%%%%%%%%%%%%
\begin{table}[htpb]
\begin{center}
\begin{tabular}{|l|l|r|}
\bf{Variable}  & \bf{Comment} & \bf{Events} \\ \hline \hline
Trigger & electron + \met & \bf{119,890}  \\
 No L\O\ requirement
in trigger &  & \bf{84,310} \\
 Main ring cuts &  & \bf{63,978} \\
  Single interaction & & \bf{17,870} \\ \hline
 \hline  One electron in fiducial region
  & $|\eta|< 1.1$ or $1.5<|\eta|< 2.5$ & \bf {17,626}   \\
 $E_T$ of electron  &   $>$ 25 GeV & \bf{15,203}  \\
Electron quality  & isolation, shape, EM fraction & \bf{13,770}
\\  $\met$  &   $>$ 25 GeV & \bf{12,622}  \\ \hline
\hline \bf{Total $W \rightarrow e\nu$ sample} &  & \bf{12,622}  \\
 Central $e$ sample & $|\eta|< 1.1$ & \bf{8,724} \\
Forward $e$ sample & $1.5<|\eta|< 2.5$ & \bf{3,898}
\\
\end{tabular}
\vspace{4mm} %agb added to pre
\caption{Central and forward $W$ boson  event selection criteria.
} \label{t:wevents}
\end{center}
\end{table}
%%%%%%%%%%%%%%%%%%%%%%%%%%%%%%%%%%%%
%\vspace{2cm}
%%%%%%%%%%%%%%%%%%%%%%%%%%%%%%%%%%%%
\begin{table}[htpb]
\begin{center}
\begin{tabular}{|l|l|r|}
\bf{Variable}  &  \bf{Comment} & \bf{Events} \\ \hline \hline
Trigger & two electrons & \bf{13,912}  \\  No L\O\ requirement in
trigger &  & \bf{10,023} \\ Main ring cuts &  & \bf{8,751} \\

  Single interaction & & \bf{2,381} \\ \hline
 \hline Two electrons in fiducial region
  & $|\eta|< 1.1$ or $1.5<|\eta|< 2.5$ & \bf
{1,617}
\\  $E_T$ of electrons   & $>$ 25 GeV & \bf{1,046}
\\
Electron quality  & isolation, shape, EM fraction & \bf{893} \\
 Invariant mass window & $76< M_{ee}< 106$ \gevcc & \bf{811}
\\ \hline \hline  \bf{Total $Z \rightarrow e^+e^-$ sample} &  & \bf{811}
\\
\end{tabular}
\vspace{4mm} %agb added to pre
\caption{ $Z$ boson  event selection criteria.  }
\label{t:zevents}
\end{center}
\end{table}
%%%%%%%%%%%%%%%%%%%%%%%%%%%%%%%%%%%%

The other analysis cuts are all standard criteria employed in D\O\
electron analyses. In addition to a $25$ GeV threshold on the
event $\met$ and the electron $E_T$, and further selection based
on the electron quality, events that occurred during the injection
of proton bunches in the Main Ring accelerator are rejected (these
often produced significant energy deposition in the D\O\
calorimeter)~\cite{wmass_prd}. The final data samples consist of
811 $Z$ boson candidate events, and 12,622 $W$ boson candidate
events, of which 8,724 have a central electron ($|\eta|< 1.1$) and
3,898 have a forward electron ($1.5<|\eta|< 2.5$). A summary of
the event selections is given in Tables~\ref{t:wevents}
and~\ref{t:zevents}.

\begin{figure}[tpb]

\epsfxsize=4.0in %4 for pre 3 for PRL
\vspace{2mm} %add to PRE
%%%agb v4 \centerline{\epsffile{paper_plots/paper2_allw_2d.eps}}
%%% for v5 new fig paper2_w2d_ncal250-slash.eps has truncation and slash,then do new bounding box
\centerline{\epsffile{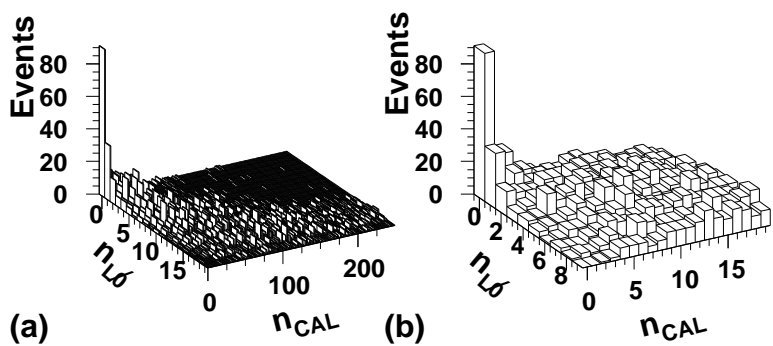}} \vspace{4mm}
    \caption{The multiplicity in forward tiles (\nl) and in calorimeter towers
     (\ncal) for the forward region with the lower multiplicity, for the
     combined central and forward electron $W$ boson samples: (a) shows the full range of multiplicity,
     (b) shows expanded region of low multiplicity ($\ncal<20$, $\nl<10$).}
    \label{f:wlego}

\end{figure}

\begin{figure}[htpb]

\epsfxsize=3.5in %4 for pre 3 for PRL
\vspace{2mm} %add to PRE
%%%agb v4 \centerline{\epsffile{paper_plots/paper_cen-fwd-z-2dmult.eps}}
%%%for v5 paper_cfz-2dmult-slash.eps adds slash copied to new name
\centerline{\epsffile{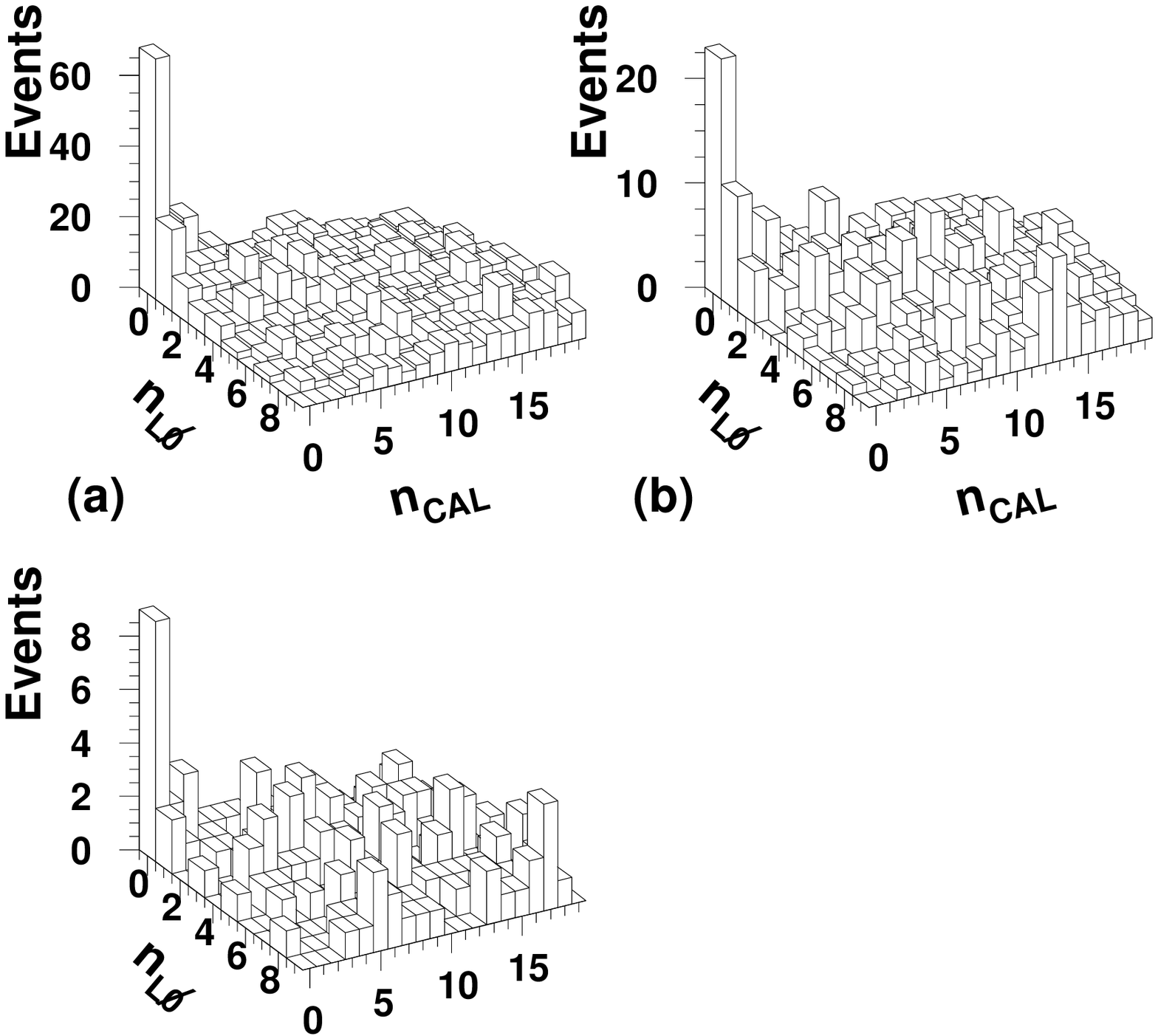}} \vspace{8mm}
    \caption{The forward tile versus calorimeter tower multiplicity  for
    (a) the central $W$ boson sample,
    (b) the forward $W$ boson  sample, and (c) the $Z$ boson  sample.}
    \label{f:3lego}

\end{figure}

Figure~\ref{f:wlego} shows two views of  \nl\ versus \ncal\ for
the combined central and forward $W$ boson sample. The
multiplicity in the forward $\eta$ interval with the lower \ncal\
multiplicity (for some events this interval is at $+\eta$ and for
others $-\eta$) is plotted for Fig.~\ref{f:wlego}(a) and (b) for
the full range of  multiplicity
%(no cutoff on \ncal\ or \nl)
and for the region of low multiplicity ($\ncal<20$, $\nl<10$),
respectively. The distributions peak at zero multiplicity
($\ncal=\nl=0$), in qualitative agreement with expectations for a
diffractive component in the data. Figure~\ref{f:3lego} shows this
scatter plot separately for the (a) central and (b) forward $W$
boson samples, and for the (c) $Z$ boson sample. All samples show
clear evidence for a diffractive component at low multiplicity.

%%Fig 1 was fig 22 2d and 1d central W
%%%%%%%%%%%%%%%%%%%%%%%%%%%%%%%%%%%%%%%%%%%%%%%%%%%%%%%%%%%%%%%%
%%%%%%%%%%%%%%%%%%%%%%%%%%%%%%%%%%%%%%%%%%%%%%%%%%%%%%%%%%%%%%%%
%%%Fig2 was fig 36, 3mult distributions
%%%%%%%%%%%%%%%%%%%%%%%%%%%%%%%%%%%%%%%%%%%%%%%%%%%%%%%%%%%%%%%%
%%%%%%%%%%%%%%%%%%%%%%%%%%%%%%%%%%%%%%%%%%%%%%%%%%%%%%%%%%%%%%%%

\begin{figure}[htpb]

\epsfxsize=3.0in %4 for pre 3 for PRL
\vspace{2mm} %add to PRE
%%%agb v4 \centerline{\epsffile{paper_plots/paper_char_w_new.eps}}
%%% V5 has new limits paper_char_w_may03.eps copied to fig3_chars.eps
\centerline{\epsffile{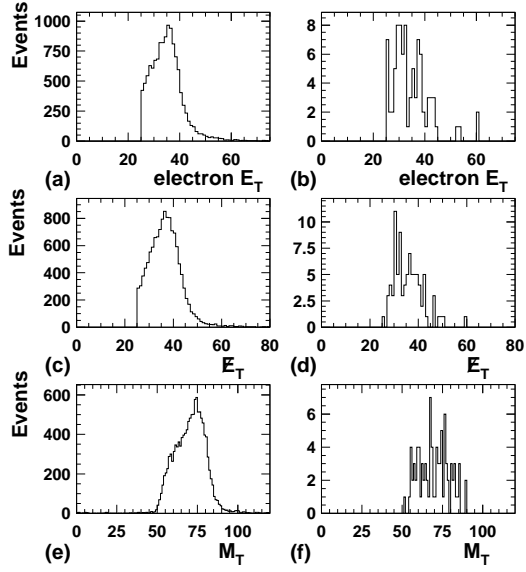}} \vspace{4mm}
    \caption{Event characteristics for standard $W$ boson events
    (left column, $\ncal > 1$), compared to diffractive $W$ boson candidates
    (right column, $\ncal=\nl=0$).  The top plots compare electron $E_T$,
    the middle plots show  \met, and the bottom plots compare the transverse mass
    ($M_T$) for the two cases.}
    \label{f:wchar}

\end{figure}

We now compare characteristics of the diffractive $W$ boson
candidates to the non-diffractive events to verify that these are
typical $W$ bosons, except for the presence of a rapidity gap.
Figures~\ref{f:wchar}(a), (c), and (e) show the electron $E_T$,
\met, and transverse mass ($M_T$), respectively,  for standard $W$
boson events ($\ncal>$ 1), while Figs.~\ref{f:wchar}(b), (d), and
(f) show the corresponding quantities for  diffractive candidate
events ($\nl=\ncal=0$). Although the statistics in the diffractive
sample are limited, the distributions in all three variables are
very similar. The mean values for these quantities for the
non-diffractive and diffractive samples, respectively, are in
excellent agreement: $<E_T>=35.2$ versus $35.1$ GeV, $<\met>=36.9$
versus 36.5 GeV, and $<M_T>=70.4$ versus 72.5 \gevcc, with
uncertainties of about 3\% on the latter values due to the limited
statistics for diffractive candidates (91 events).
%%%Fig4 was Fig 34 shows W event chars
%%%%%%%%%%%%%%%%%%%%%%%%%%%%%%%%%%%%%%%%%%%%%%%%%%%%%%%%%%%%%%%%
%%%%%%%%%%%%%%%%%%%%%%%%%%%%%%%%%%%%%%%%%%%%%%%%%%%%%%%%%%%%%%%%

\begin{figure}[htpb]
\vspace{-0.5in}
  \epsfxsize=4.0in %4 for pre 3 for PRL
\vspace{2mm} %add to PRE
\centerline{\epsffile{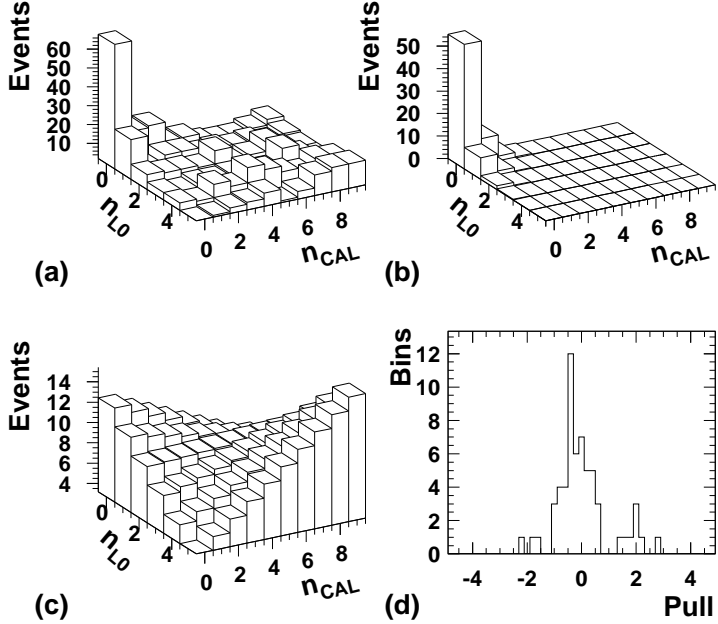}} \vspace{6mm}

    \caption{The (a) \nl\ versus \ncal\ distribution for the central electron $W$ boson data
    from Fig.~\ref{f:3lego}(a), and corresponding
(b) fitted signal, (c) fitted background, and (d) normalized pull
distributions.}
    \label{f:wfit}

\end{figure}

The fractions of $W$ and $Z$ boson events that contain forward
rapidity gaps (``gap fraction'') are extracted from fits to the
data in Fig.~\ref{f:3lego}. The non-diffractive (high
multiplicity) background in the signal region is represented by a
four-parameter polynomial surface, and the signal by a
two-dimensional falling exponential as in Ref.~\cite{D0jetPLB}.
%, as suggested by Monte Carlo~\cite{Kristal}.
Figure~\ref{f:wfit} shows the multiplicity distribution from
Fig.~\ref{f:3lego}(a), and the resulting fitted signal, fitted
background, and the normalized distribution of pulls
([data-fit]/$\sqrt{N}$). The $\chi^2/{\rm dof}=1.04$ for this fit,
and all other fits are of comparable quality.

%%AGB careful that -0.5 doesn't mess this up, okay for now
%%%%%%%%%%%%%%%%%%%%%%%%%%%%%%%%%%%%%%%%%%%%%%%%%%%%%%%%%%%%%%%%
%%%%%%%%%%%%%%%%%%%%%%%%%%%%%%%%%%%%%%%%%%%%%%%%%%%%%%%%%%%%%%%%

The fitting process is repeated over a systematically varied range
of \nl\ (lower limit of 2 tiles and upper limit ranging from 5 to
7 tiles) and \ncal\ (lower limit of 3 towers and upper limit
ranging from 6 to 12 towers) to minimize any dependence of the
signal on the chosen region. The result is a distribution of gap
fractions, with the final signal defined by the mean of this
distribution. The statistics in the $Z$ boson sample are
insufficient to perform an independent fit, so we use the
background shape from the combined $W$ boson sample scaled to the
$Z$ boson data to determine the diffractive $Z$ boson signal.
Varying the shape of the background samples used for the fit shows
only small variations in the signal, well within the quoted
uncertainties.

To determine the final gap fractions, we correct the fitted values
for residual contamination from  multiple interaction events which
were not rejected by the single interaction requirement. These
events contribute only to the denominator of the gap fraction,
resulting in a measured gap fraction that is lower than the
correct value. This correction is determined using the full data
sample with no single interaction requirement by comparing the
predicted number of single interaction events (based on the
instantaneous luminosity) with the observed number of events after
the single interaction selection. This  method demonstrates that
our single interaction requirement is quite effective, and yields
only an absolute correction of $(0.09 \pm 0.05)\%$ for the central
electron $W$ boson and the $Z$ boson samples and a negligible
correction for the forward electron $W$ boson sample.

Table~\ref{t:gapfrac} summarizes the final gap fractions obtained
for the $W$ and $Z$ boson samples and their significances. The
combined $W$ boson sample has a gap fraction of
$0.89\pm^{0.19}_{0.17}\%$ and a probability that the
non-diffractive background fluctuated to the diffractive signal of
$3 \times 10^{-14}$, corresponding to a  significance of
7.5$\sigma$. The central $W$ boson fraction (electron
$|\eta|<1.1$) of  $1.08 \pm^{0.19}_{0.17}\%$ is greater than the
forward fraction ($1.5<|\eta|< 2.5$) of $0.64
\pm^{0.18}_{0.16}\%$, unlike for jet events~\cite{D0jetPLB} (or
typical diffractive expectations), which have a larger forward
fraction. The $Z$ boson sample has a gap fraction of $1.44
\pm^{0.61}_{0.52}\%$, with a significance of 4.4$\sigma$.
Uncertainties are dominated by those on the fit parameters.
Additional small uncertainties from the dependence on the range of
multiplicities used in the fits are added in quadrature. Potential
sources of systematic error, such as the number of fit parameters,
electron quality criteria, tower thresholds, and residual noise,
yield only negligible variations in the gap
fractions~\cite{Kristal,coney}.
%%%%results table
%%%%%%%%%%%%%%%%%%%%%%%%%%%%%%%%%%%%%%%%%%%%%%%%%%%%%%%%%%%%%%%%
\begin{table}[htpb]
\begin{center}
\begin{tabular}{|l|l|l|}
\bf{Sample}  & \bf{Gap Fraction} & \bf{ Significance }
\\ \hline \hline

Central $W \rightarrow e\nu $ ($|\eta |< 1.1$) & \bf{($1.08 + 0.19
- 0.17)\%$} & $1 \times 10^{-14}$ (7.7$\sigma$) \\

Forward $W \rightarrow e\nu $ ($1.5<|\eta|< 2.5$) & \bf{$(0.64 +
0.18 - 0.16)\%$} & $6 \times 10^{-8}$ \,(5.3$\sigma$) \\ Total $W
\rightarrow e\nu $  & \bf{$(0.89 + 0.19 - 0.17)\%$} &
  $3 \times 10^{-14}$ (7.5$\sigma$) \\
 Total $Z \rightarrow e^+ e^- $   & \bf{$(1.44 + 0.61 -
0.52)\%$} & $5 \times 10^{-6}$ \, (4.4$\sigma$) \\
\end{tabular}
\vspace{4mm} %agb added to pre
\caption{  Measured gap fractions and probabilities for
non-diffractive $W$ and $Z$ boson events to fluctuate and mimic
diffractive $W$ and $Z$ boson production.} \label{t:gapfrac}
\end{center}
\end{table}
%%%%%%%%%%%%%%%%%%%%%%%%%%%%%%%%%%%%%%%%%%%%%%%%%%%%%%%%%%%%%%%%

We have thus far considered only non-diffractive $W$ and $Z$ boson
events as the relevant background to diffractive production. We
now consider contamination from events other than the desired $W
\rightarrow e\nu$ and $Z \rightarrow ee$ states, drawing primarily
on the work of Ref.~\cite{wwidth_prd}.

The largest background to $W$ boson production is from multijet
events in which one jet is misidentified as an electron, while
another is measured incorrectly, thereby providing  large \met.
The fraction of fake $W \rightarrow e\nu$ events from multijet
production was calculated~\cite{wwidth_prd} to be $0.046 \pm
0.014$ and $0.143 \pm 0.043$ of the total $W \rightarrow e\nu$
events for the central and forward electron samples, respectively.
We use these fractions to determine the number of multijet events
in our samples, and use measurements from Ref.~\cite{D0jetPLB} to
obtain the number of diffractive events expected from this
background. The results are shown in Table~\ref{w_qcd_background}:
for the central electron sample, a total of 0.88 diffractive
events are expected from the 401 multijet background events. Given
that there are 8724 events in the central electron $W$ boson
sample, with a measured diffractive fraction of $1.08
\pm^{0.19}_{0.17}\%$, we expect a total of $ 94 \pm^{17}_{15}$
diffractive events. Recalculating the central $W$ boson gap
fraction after subtracting the multijet background gives a
slightly higher value of $1.11\%$ (93/8323). We note that this 3\%
change is an upper limit, because multijet background in events
with single interactions would likely be smaller than in the
inclusive $W$ boson sample due to smaller fluctuations expected in
$\met$. The forward sample gives a negligible correction, since
the gap fraction from diffractive $W$ boson signal and multijet
background are nearly identical.
%%%%%%%%%%%%%%%%%%%%%%%%%%%%%%%%%%%%%%%%%%%%%%%%%%%%%%%%%%%%%%%%
\begin{table}[htpb]
\begin{center}
\begin{tabular}{|c|c|c|c|c|c|}
Sample& Total & Multijet & Multijet & Diffractive & Diffractive

\\
      & Events & Fraction   & Events  & Dijet Fraction & Multijet Events \\
      \hline\hline
 \bf{central $W$} & 8724 & $0.046 \pm 0.014$ & 401 &
$0.22 \pm 0.05\%$ & 0.88 \\  \bf{forward $W$} & 3898 & $0.143 \pm
0.043$ & 557 & $0.65\ \pm 0.04\%$ & 3.6 \\
\end{tabular}
\vspace{4mm} %agb added to pre
\caption{  The number of multijet background events in the
diffractive $W$ boson sample is calculated. Then, given the number
of multijet events in the sample and the diffractive dijet rate,
the number of diffractive events expected from these background
events is calculated. } \label{w_qcd_background}
\end{center}
\end{table}
%%%%%%%%%%%%%%%%%%%%%%%%%%%%%%%%%%%%%%%%%%%%%%%%%%%%%%%%%%%%%%%%

In addition to the multijet background, we consider background
from misidentified $Z$ boson events in which one electron is not
detected. Again using methods from Ref.~\cite{wwidth_prd}, we
estimate $94 \pm 24$ $Z$ boson events in the combined $W$ boson
sample, with 1.35 of these being diffractive $W$ boson candidates.
Subtracting this background would result in a less than 1\%
correction, in the opposite direction from the multijet
correction, since the diffractive $Z$ boson signal is larger than
that of the diffractive $W$ boson. Finally, the background level
from $W \rightarrow \tau\nu$ is expected to be small (about 2\%),
and we would expect the same gap fraction from $W$ bosons that
decay to $\tau$ leptons as from the electron channel, therefore no
correction is needed.

Combining all these background sources yields a total background
to diffractive $W$ boson production of at most 2\%, which is
insignificant compared to the total 20\% uncertainty, and we
therefore do not apply any correction. Similar considerations for
the diffractive $Z$ boson sample yield at most a 4\% background
correction factor, which is again not significant, and
consequently not applied.

In this paper we have chosen to present the gap fraction, which is
directly based on observable quantities. We thus avoid the
reliance on potentially large model-dependent corrections.
Therefore the measured gap fraction of $1.08 \pm^{0.19}_{0.17}\%$
for central electron $W$ boson events cannot be directly compared
to the CDF measurement of $(1.15 \pm 0.55)$\%~\cite{CDFW2}, which
includes a correction factor derived from the {\footnotesize
POMPYT} diffractive Monte Carlo~\cite{POMPYT} (based on the
Ingelman-Schlein model) to attempt to account for how often a
diffractive event does not yield a rapidity gap. They obtained a
correction factor of 0.81 based on their Beam-Beam Counter (BBC)
multiplicity, implying an uncorrected value of $(0.93 \pm
0.44)$\%, which is consistent with our measurement. However, this
correction factor is quite different from that obtained by D\O\
and CDF using two-dimensional (luminosity counter and calorimeter
multiplicity) methods subsequently adopted by both collaborations
to extract rapidity gap signals. We obtain a correction factor of
$0.21\pm 0.04$ from our diffractive $W$ boson Monte Carlo using a
quark structure for the pomeron, which compares well with the
quark-structure-based correction for our central jet measurement
(0.18)~\cite{Kristal} and the CDF correction factor for their
diffractive $b$-quark production (0.22)~\cite{CDFb} and J/$\Psi$
production (0.29)~\cite{CDFjpsi}. Since there is no consensus on
the correct model for describing diffractive data at the Tevatron,
we feel that using {\footnotesize POMPYT} to correct the data is
not advisable, but it should be noted that our corrected $W$ boson
gap fraction would be 5.1\% according to this model, while there
would be no correction needed for non-pomeron models such as SCI.

Next, we calculate the ratio of the diffractive $W$ and $Z$ boson
cross sections. In addition to intrinsic interest in this
measurement, it is a potentially important input to the systematic
uncertainty on the ratio $R$ of the two cross
sections~\cite{wwidth_prd}. We can write the diffractive cross
section ratio $R_D$ in terms of the gap fractions and the ratio
$R$ as follows:

%%%%%%%%%%%%%%%%%%%
\begin{equation}
R_D=\frac{W_D}{Z_D} = (\frac{W_D}{W}/ \frac{Z_D}{Z}) \times R =
6.45\pm^{3.06}_{2.64},
\end{equation}

\noindent where we have substituted the measured gap fractions
$W_D/W$ and $Z_D/Z$ from this Letter, and the measured value $R =
10.43 \pm 0.15 (stat) \pm 0.20 (syst) \pm 0.10
(NLO)$~\cite{wwidth_prd}, which takes into account acceptance
differences between the $W$ and $Z$ boson samples (assumed to be
similar for diffractive and non-diffractive events). This value of
$R_D$ is consistent with the ratio for non-diffractive production.

\begin{figure}[htpb]
 \epsfxsize=3.5in %4 for pre 3 for PRL
\vspace{2mm} %add to PRE
\centerline{\epsffile{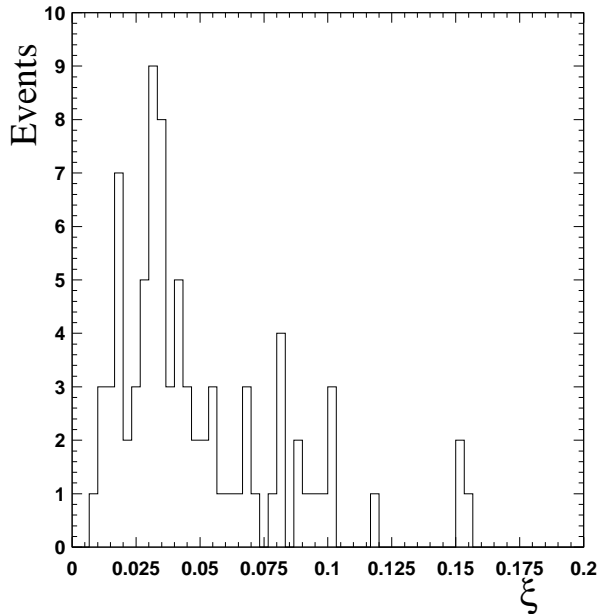}} \vspace{4mm}
    \caption{The $\xi$ distribution for the combined central and forward electron
    $W$ boson data with $\ncal=\nl=0$, extracted from calorimeter information
     as described in the text.}
    \label{fig:wdata_xi}

\end{figure}

Finally, we measure the fractional momentum loss of the scattered
proton $\xi$ using the following equation~\cite{collins}:

\begin{equation}
\xi \approx \frac{1}{\sqrt{s}}\sum_{i} E_{T_i}e^{\eta_i} \\
\label{eq:xi}
\end{equation}

\noindent where $E_{T_i}$ and $\eta_i$ denote the transverse
energy and pseudorapidity, respectively, of the observed
particles. The $\eta$ of the outgoing scattered proton or
antiproton (and the rapidity gap) is defined to be positive. As
discussed in Ref.~\cite{D0jetPLB}, Eq.~\ref{eq:xi} is particularly
sensitive to particles emitted in the well-measured central region
near the rapidity gap, while particles lost down the beam pipe at
negative $\eta$ have negligible effect. Using a sample of
{\footnotesize POMPYT} $W$ boson events, where $\xi$ can be
determined from the momentum of the scattered proton, we have
verified that Eq.~\ref{eq:xi} is valid independent of pomeron
structure. A scale factor $1.5 \pm 0.3$, derived by passing the
Monte Carlo data through a full detector simulation, is used to
convert $\xi$ measured from all particles to that from just the
electromagnetic energy depositions in the
calorimeter~\cite{coney}. Figure~\ref{fig:wdata_xi} shows the
$\xi$ distribution for the diffractive $W$ boson candidate event
sample with $\ncal=\nl=0$. The mean $\xi$ is 0.052 and most of the
events have $\xi<0.1$. Comparison of this $\xi$ distribution
obtained from calorimeter information with that from the measured
proton using the upgraded D\O\ detector in Run II will give
important insight into the nature of diffraction.

%Fig 5 was figure 57 from MC section
%%%%%%%%%%%%%%%%%%%%%%%%%%%%%%%%%%%%%%%%%%%%%%%%%%%%%%%%%%%%%%%%
%%%%%%%%%%%%%%%%%%%%%%%%%%%%%%%%%%%%%%%%%%%%%%%%%%%%%%%%%%%%%%%%

In conclusion, we have observed diffractive $W$ boson production
with greater than 7$\sigma$ significance, shown that these
diffractive $W$ boson candidates  have similar properties to
non-diffractive ones, and measured the fraction of $W$ boson
events that are produced diffractively, in both the central and
forward regions. We also have provided the first evidence for
diffractive $Z$ boson production. The extracted gap fractions have
no model-dependent corrections, and are typically about 1\%, far
below expectations for a quark-dominated pomeron. We have obtained
a ratio of diffractive $W$ and $Z$ boson cross sections consistent
with the ratio for non-diffractive production. We have also
measured the fractional momentum lost by the scattered proton and
found that typically less than 10\% of the proton momentum takes
part in the hard scattering, with an average of about 5\%.

% acknowledgement_paragraph_r1.tex
% acknowledgement_paragraph_r1.tex            5/30/03
%
We thank the staffs at Fermilab and collaborating institutions, 
and acknowledge support from the 
Department of Energy and National Science Foundation (USA),  
Commissariat  \` a L'Energie Atomique and 
CNRS/Institut National de Physique Nucl\'eaire et 
de Physique des Particules (France), 
Ministry for Science and Technology and Ministry for Atomic 
   Energy (Russia),
CAPES, CNPq and FAPERJ (Brazil),
Departments of Atomic Energy and Science and Education (India),
Colciencias (Colombia),
CONACyT (Mexico),
Ministry of Education and KOSEF (Korea),
CONICET and UBACyT (Argentina),
The Foundation for Fundamental Research on Matter (The Netherlands),
PPARC (United Kingdom),
Ministry of Education (Czech Republic),
A.P.~Sloan Foundation,
and the Research Corporation.
%

%
%pre add
\clearpage

%remove vspace for pre
%\vspace{-0.5cm}


\begin{thebibliography}{99}
%\vspace{-1.3cm}
% list_of_visitor_addresses_r1.tex                            8/31/01
%
%\bibitem[\dag]{name}
\bibitem[*]{lehner}
Visitor from University of Zurich, Zurich, Switzerland.
\bibitem[\dag]{przybycien}
Visitor from Institute of Nuclear Physics, Krakow, Poland.
%
\vskip 0.25cm


\bibitem{regge} P.D.B. Collins, {\it
An Introduction to Regge Theory and High Energy Physics},
Cambridge University Press, Cambridge (1977).

\bibitem{IS}   G. Ingelman and P. Schlein, Phys. Lett. B {\bf 152}, 256
 (1985).




%agb check if sci published

\bibitem{UA8} R. Bonino {\it et al.} (UA8 Collaboration),
Phys. Lett. B {\bf 211}, 239 (1988); A. Brandt {\it et al}. (UA8
Collaboration), Phys. Lett. B {\bf 297}, 417 (1992).

\bibitem{CDFJ}  F. Abe {\it et al.} (CDF Collaboration),
                 Phys. Rev. Lett. {\bf 79}, 2636 (1997).
\bibitem{Zeus}  J. Breitweg {\it et al.} (ZEUS Collaboration), Eur. Phys. J.
{\bf C5}, 41 (1998) and references therein.

\bibitem{H1}  C. Adloff {\it et al.} (H1 Collaboration),
              Eur. Phys. J. {\bf C6}, 421 (1999).
\bibitem{CDFb} T. Affolder {\it et al.} (CDF Collaboration),
                 Phys. Rev. Lett. {\bf 84}, 232 (2000).

\bibitem{CDFjpsi} T. Affolder {\it et al.} (CDF Collaboration),
                 Phys. Rev. Lett. {\bf 87}, 241802 (2001).

\bibitem{fact}  L. Alvero, J.C. Collins, J. Terron and J. Whitmore, Phys.
Rev. D {\bf 59}, 74022 (1999).


\bibitem{CDFpbj} T. Affolder {\it et al.} (CDF Collaboration),
                 Phys. Rev. Lett. {\bf 84}, 5043 (2000).

\bibitem{D0jetPLB} B. Abbott  {\it et al.} (D\O\ Collaboration),
 Phys. Lett. B {\bf 531}, 52 (2002).

\bibitem{khoze} V. A. Khoze, A. D. Martin, and M.G. Ryskin,
 Phys. Lett. B {\bf 502}, 87 (2001) and references therein.

\bibitem{SCI} A. Edin, G. Ingelman, and J. Rathsman, J. Phys. G
              {\bf 22}, 943 (1996).

\bibitem{bruni} P. Bruni and G. Ingelman, Phys. Lett.  B {\bf 311}, 317 (1993).

\bibitem{SCI2}  N. Timneanu, R. Enberg,
              and G. Ingelman,  hep-ph/0111210, 2001
              (unpublished).

\bibitem{CDFW2}  F. Abe {\it et al.} (CDF Collaboration),
                 Phys. Rev. Lett. {\bf 78}, 2698 (1997).


\bibitem{NIM}  S. Abachi {\it et al.} (D\O\ Collaboration),
              Nucl. Instrum. Methods in Phys. Res. A {\bf 338}, 185 (1994).

\bibitem{Kristal} K.~Mauritz,
 Ph.D.~Dissertation, Iowa State University, 1999 (unpublished).
%agb remove jets
%\bibitem{D0jets} B. Abbott {\it et al.} (D\O\ Collaboration),
%          Nucl. Instrum. Methods Phys. Res. A {\bf 424}, 352 (1999).

%\bibitem{prdjet} B. Abbott {\it et al.} (D\O\ Collaboration),
% Phys. Rev. D {\bf 64}, 12004 (2001).

%\bibitem{PYTHIA}
%H.-U. Bengtsson and T. Sj\"ostrand, Comp.
%              Phys. Comm. {\bf 46}, 43 (1987); T. Sj\"ostrand, CERN-TH.6488/92.
%

%\bibitem{DL}A. Donnachie and P.V. Landshoff, Nucl. Phys. B {\bf 303},
%634 (1988).


%bibitem{renorm} K. Goulianos, Phys. Lett. B {\bf 358}, 379
%(1995).

\bibitem{wmass_prd} B. Abbott et al. (D\O\ Collaboration), Phys. Rev. D {\bf 58},
    092003 (1998).
\bibitem{wwidth_prd} B. Abbott et al. (D\O\ Collaboration), Phys. Rev. D {\bf 61},
    072001 (2000).

\bibitem{coney} L. Coney, Ph. D. Thesis: Diffractive $W$ and $Z$ Boson
Production in \pbarp\ Collisions at $\rs=1800$ GeV, 2000
(unpublished).


\bibitem{POMPYT}
 P. Bruni and G. Ingelman, DESY 93-187, 1993 (unpublished). We used
a modified version of  {\footnotesize POMPYT} 2.6.

\bibitem{collins} J. Collins, hep-ph/9705393, 1997 (unpublished).


\end{thebibliography}
\end{document}